\documentclass[conference,a4paper,10pt]{IEEEtran}
\usepackage{cite}
\usepackage[pdftex]{graphicx}
\pdfoutput=1
\usepackage {times}
\usepackage{graphicx}
\usepackage{color}
\usepackage{amsmath}
\usepackage{amstext}
\usepackage{hyperref}
\usepackage{amsfonts}
\usepackage{amssymb}
\usepackage{psfrag}
\usepackage{fancyhdr}
\usepackage[plain]{fancyref}
\usepackage{dsfont}
\usepackage{bbm}
\usepackage{blkarray}
\usepackage{subfigure}
\usepackage{epsfig}
\usepackage[ruled]{algorithm}
\usepackage{algpseudocode}
\usepackage{tabularx}
\newcommand*{\fancyrefalgolabelprefix}{algo}
\frefformat{plain}{\fancyrefalgolabelprefix}{algorithm~#1}
\Frefformat{plain}{\fancyrefalgolabelprefix}{Algorithm~#1}
\Frefformat{plain}{\fancyrefeqlabelprefix}{(#1)}
\Frefformat{plain}{\fancyreftablabelprefix}{Table~#1}
\usepackage{etoolbox}
\let\bbordermatrix\bordermatrix
\patchcmd{\bbordermatrix}{8.75}{4.75}{}{}
\patchcmd{\bbordermatrix}{\left(}{\left[}{}{}
\patchcmd{\bbordermatrix}{\right)}{\right]}{}{}

\DeclareGraphicsExtensions{.pdf}
\IEEEoverridecommandlockouts

\begin{document}

\title{Dynamic Uplink-Downlink Optimization in TDD-based Small Cell Networks}

\author{
\IEEEauthorblockN{Mohammed S. ElBamby\IEEEauthorrefmark{1}, Mehdi Bennis\IEEEauthorrefmark{1}, Walid Saad\IEEEauthorrefmark{2} and Matti Latva-aho\IEEEauthorrefmark{1} \\}
\IEEEauthorblockA{\small\IEEEauthorrefmark{1}Centre for Wireless Communications, University of Oulu, Finland, \\ email: \{melbamby,bennis,matti.latva-aho\}@ee.oulu.fi \\
\IEEEauthorrefmark{2}Wireless@VT, Bradley Department of Electrical and Computer Engineering, Virginia Tech, Blacksburg, VA, USA, email: walids@vt.edu}
\thanks{This research is supported by the SHARING project under Finland grant 128010 and the U.S. National Science Foundation (NSF) under Grants CNS-1253731 and CNS-1406947.}
}

\IEEEpubid{\makebox[\columnwidth]{978-1-4799-5863-4/14/\$31.00~\copyright~2014 IEEE \hfill}\hspace{\columnsep}\makebox[\columnwidth]{}}

\maketitle

\begin{abstract}
Dynamic Time-division duplex (TDD) can provide efficient and flexible splitting of the common wireless cellular resources between uplink (UL) and downlink (DL) users. In this paper, the UL/DL optimization problem is formulated as a noncooperative game among the small cell base stations (SCBSs) in which each base station aims at minimizing its total UL and DL flow delays. To solve this game, a self-organizing UL/DL resource configuration scheme for TDD-based small cell networks is proposed. Using the proposed scheme, an SCBS is able to estimate and learn the UL and DL loads autonomously while optimizing its UL/DL configuration accordingly. Simulations results show that the proposed algorithm achieves significant gains in terms of packet throughput in case of asymmetric UL and DL traffic loads. This gain  increases as the traffic asymmetry increases, reaching up to $97\%$ and $200\%$ gains relative to random and fixed duplexing schemes respectively. Our results also show that the proposed algorithm is well-
adapted to dynamic traffic conditions and different network sizes, and operates
efficiently in case of severe cross-link interference in which neighboring cells transmit in opposite directions.\\ \\
\emph{Keywords- Dynamic-TDD; small cells; reinforcement learning; self-organizing networks}

\end{abstract}

\IEEEpeerreviewmaketitle

\section{Introduction}

\begin{figure}
\centering
\includegraphics[width = 3.25in, trim=0cm 0cm 11.5cm 24cm, clip=true]{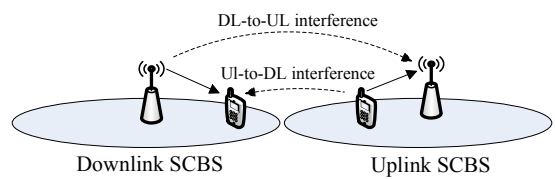}
\caption{New types of interference arising in the dynamic TDD configuration.}
\label{fig:interference_types}
\end{figure}

Next generation of heterogeneous networks (HetNets) are expected to have significant variations in traffic load between different cells and at different times. Moreover, due to the massive increase in the use of smartphones and video streaming applications \cite{ericsson_rep}, an asymmetric and dynamically changing uplink (UL) and downlink (DL) traffic is expected, incurring new types of cross-link interferences. In order to cope with this challenge, it is necessary that the evolution of current wireless networks is able to accommodate asymmetric UL and DL traffic loads. While Time-division duplex (TDD) systems \cite{LTE_book} have the capability of handling this asymmetry, in practice, cells operating in TDD are assumed to synchronize their UL and DL transmissions. Otherwise, they can suffer from a new type of interference from base stations transmitting in the opposite direction; this is referred to as \emph{cross-link interference} \cite{LTE_book}.  Basically, there are two types of interference associated with the asynchronous TDD operation, UL-to-DL interference and DL-to-UL interference, as illustrated in \Fref{fig:interference_types}.

Dealing with cross-link interference is a key challenge for deploying dynamic TDD-based systems \cite{dynamic_TDD}. For example, time-division Long Term Evolution (TD-LTE) systems require synchronization between base stations over an overlapping coverage area \cite{TD_LTE_3gpp}. Moreover, under small cell base station (SCBS) deployment, this becomes more challenging \cite{guvenc_book}, since cells are more likely to have strong interference coupling, especially in dense overlapping areas. Besides, as centralized operation becomes difficult, small cells should self-organize to select their optimum UL/DL configuration, as a function of the interference levels, and users' quality-of-service (QoS) requirements.

The problem of dynamic UL-DL configuration in TD-LTE systems is studied in \cite{TDD_mag}. Therein, the prospects of dynamic TDD in TD-LTE systems are discussed and challenges are identified, among which interference management is seen as a major impediment. The performance of dynamic adaptation of UL and DL in LTE picocell systems is also analyzed in \cite{khorayev_paper}. It is shown that significant gains in packet throughput are achieved by using dynamic TDD over the synchronous TD-LTE. The analytical modeling of UL and DL performance under dynamic TDD is studied in \cite{LTE_B_paper} using tools from stochastic geometry. A cooperative UL-DL adaptation scheme is introduced in \cite{two_cells_paper} in which two SCBSs serving one user each, adapt their UL/DL configuration locally relying on exchanging low-rate information.

The main contribution of this paper is to propose a dynamic UL/DL configuration scheme for TDD-based small cell environments in which the objective is to minimize the overall UL and DL delay in each cell and in a completely autonomous manner. In particular, the problem is formulated as a noncooprative game in which the SCBSs are the players. In this game, each SCBS learns and estimates its current uplink and downlink delay, as a function of its traffic load, interference levels and flow-level dynamics, relying only on its instantaneous observations, and uses this estimated value to update its UL/DL switching point. A decentralized and self-organizing learning algorithm is then proposed to find an equilibrium of the game. The proposed approach is then shown to achieve considerable gains over fixed and random TDD deployments for different network sizes. In addition, our results show that the proposed algorithm is well-suited to small cell environments with large traffic dynamics. Simulation results show
significant packet throughput gains in case of asymmetric UL and DL traffic loads for the proposed algorithm. The gain further increases as the traffic asymmetry increases, reaching up to $97\%$ and $200\%$ gains relative to random and fixed schemes respectively.

The rest of this paper is organized as follows, in \Fref{sec:sys_form}, we describe the network model and formulate the problem. \Fref{sec:self_org} introduces the proposed self-organizing UL/DL optimization scheme. Simulation results are provided in \Fref{sec:sim_results}. Finally, \Fref{sec:conc} concludes the paper.
\section{System Model}
\label{sec:sys_form}
\subsection{Network Model}
We consider a wireless communication system consisting of a set of small cell base stations (SCBSs) $\mathcal{B} =\{1,\ldots,B\}$. We assume that a user equipment (UE) arrives at location $x$ within the considered geographical area according to a Poisson arrival process with rate $\lambda (x)$. Each UE requests either a DL or UL file whose size follows an exponential distribution with mean $1/\mu (x)$. A closed-access policy is assumed in this work, meaning that each SCBS has its own subscribed UEs, and hence no handover is considered \cite{guvenc_book}. We further assume $\mathcal{L}_b$ to be the coverage area of an SCBS $b$, where a UE at location $x$ is served by an SCBS $b$ if $x\in\mathcal{L}_b$.
\begin{figure}
\centering
\includegraphics[trim=0cm 0.60cm 0cm 0cm, clip=true]{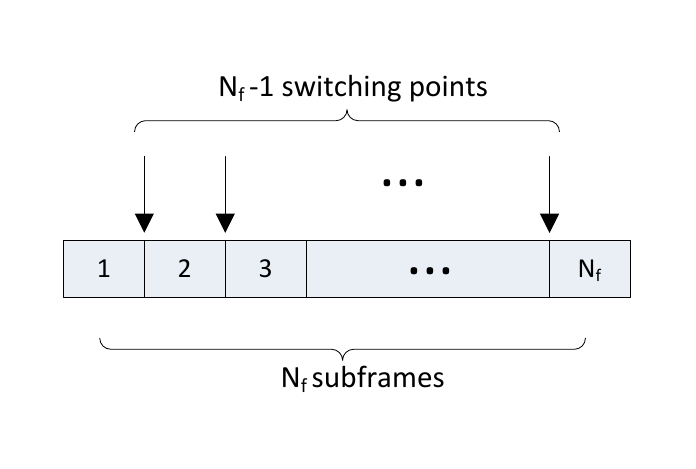}
\caption{Possible switching points in a TDD frame.}
\label{fig:TDD_frame}
\end{figure}

We assume that the system operates in TDD mode. A time frame consists of a number of $N_f$ subframes. A frame is divided into two portions, UL portion and DL portion. Each portion consists of a group of subframes dedicated to serving either UL or DL traffic. A switching point $w_b$ is defined as the point in which SCBS $b$ switches from UL mode to DL mode. There is a number of $N_f-1$ possible switching points for a frame length of $N_f$, then $w_b\in \{ 1,\ldots,W_f \}$, where $W_f = N_f-1$, as illustrated in \Fref{fig:TDD_frame}. For any of these possible switching points, there will be at least one subframe for UL and for DL in each frame.

\subsection{Problem Formulation}\label{sec:prob_form}
We define the vector  $ \boldsymbol{w} = [w_1 , w_2,\ldots, w_B]$ as the vector of switching points for all SCBSs in the system. Varying switching points asynchronously in different cells may cause opposite transmission directions in different cells which leads to cross-link interference (i.e. UL-to-DL interference and DL-to-UL interference). Consequently, the Signal-to-Interference-plus-Noise-Ratio (SINR) for UL and DL, respectively, for a receiving node at location $x\in \mathcal{L}_b$ is given by:
\begin{multline}\label{eq:SINR_UL}
\Gamma_{b}^{\text{UL}} (x) =\\
\frac{p_{b}^{\text{UL}}h_{b,b}(x) }{\sigma^2 + \sum_{j\in\mathcal{B}_\text{UL}\backslash \{b\}}p_{j}^{\text{UL}}h_{j,b}(x) + \sum_{k\in\mathcal{B}_\text{DL}}p_{k}^{\text{DL}}h_{k,b}(x) },
\end{multline}

\begin{multline}\label{eq:SINR_DL}
\Gamma_{b}^{\text{DL}} (x) =\\
\frac{p_{b}^{\text{DL}}h_{b,b}(x) }{\sigma^2 + \sum_{j\in\mathcal{B}_\text{UL}}p_{j}^{\text{UL}}h_{j,b}(x) + \sum_{k\in\mathcal{B}_\text{DL}\backslash \{b\}}p_{k}^{\text{DL}}h_{k,b}(x) } ,
\end{multline}
where $p_{b}^{\text{UL}}$ ($p_{b}^{\text{DL}}$) is the UL (DL) power from the serving node~\texttt{$b$}, $p_{j}^{\text{UL}}$ ($p_{j}^{\text{DL}}$) is the UL (DL) power from the interfering node~\texttt{$j$}, $h_{m,b}(x)$ is the channel gain, including pathloss, between the transmitting node in SCBS $b$ and the receiving node in location $x\in \mathcal{L}_b$, $\mathcal{B}_\text{UL}$ and $\mathcal{B}_\text{DL}$ are the sets of cells operating in UL and DL, respectively, and $\sigma^2$ is the noise variance. Furthermore, the data rates of a UE at location  $x\in \mathcal{L}_b$ for UL and DL, respectively, are given by:
\begin{equation}\label{eq:SE_UL}
c_{b}^{\text{UL}}(x)=f_b\log_2(1+\Gamma_{b}^{\text{UL}} (x)),
\end{equation}
\begin{equation}\label{eq:SE_DL}
c_{b}^{\text{DL}}(x)=f_b\log_2(1+\Gamma_{b}^{\text{DL}}(x)).
\end{equation}
where $f_b$ is the bandwidth allocated to that UE.

The system-load density at location $x$ is defined as \cite{load_journal}:
\begin{equation}\label{eq:sys_load}
    \varrho_b^{(l)}(x) :=  \frac{\gamma^{(l)}(x)}{c_b^{(l)}(x)},
\end{equation}
where $l\in\{\text{UL},\text{DL}\}$ and  $ \gamma^{(l)}(x) := \lambda^{(l)}(x) / \mu^{(l)}(x) $ is the load density at location $x$.

The \emph{cell load density} for cell $b \in \{1,2,\ldots,B\}$ is defined as the time delay needed to serve the UL and DL traffic as follows:
\begin{equation}\label{eq:load}
    \rho_b^{(l)}(w_b) =  \frac{1}{\delta^{(l)}(w_b)}\int_{x\in \mathcal{L}_b} \varrho_b^{(l)}(x)dx.
\end{equation}
where $l\in\{\text{UL},\text{DL}\}$, $\delta^{(l)}(w_b)$ is the UL or DL duty cycle, which is the fraction of time frames dedicated to either UL or DL service within a frame, and is expressed as follows:
\begin{equation}\label{delta_eq}
  \delta^{(l)}(w_b)=
\begin{cases}
   \dfrac{w_b}{W_f}& l=\text{UL},\\
   \\
   \dfrac{W_f-w_b}{W_f}& l=\text{DL}.
\end{cases}
\end{equation}

Here, dividing each cell load by its respective UL or DL duration is done in order to account for the UL/DL effective traffic. Therefore, lower duty cycles lead to higher delays and vice versa. Our objective is to find the vector of switching points  $\boldsymbol{w}$ that minimizes the overall average flow delay by minimizing $\sum_{b \in \mathcal{B}} \frac{\rho_b}{1-\rho_b}$ over the entire time frame \cite{load_journal}. Therefore, we define a cost function that reflects the flow delay average over the whole subframes within a timeframe, calculated as follows:
\begin{multline}\label{eq:cost_fun}
J(\boldsymbol{w}) = \sum_{b=1}^{B} \Biggr( \frac{1}{w_b}\sum_{j=1}^{w_b} \frac{\rho_{b,j}^{\text{UL}}(w_b)}{1-\rho_{b,j}^{\text{UL}}(w_b)} \\
+  \frac{1}{W_f-w_b}\sum_{j=w_b+1}^{W_f} \frac{\rho_{b,j}^{\text{DL}}(w_b)}{1-\rho_{b,j}^{\text{DL}}(w_b)}\Biggl)
\end{multline}
where $ \rho_{b,j}^{\text{UL}}(w_b)$ and $ \rho_{b,j}^{\text{DL}}(w_b)$ are the $b$-th cell load densities for the UL and DL, respectively, at subframe $j$ as defined in \Fref{eq:load}. Thus, we can define the following cost optimization problem:
\begin{align}\label{eq:cost_prob}
& \underset{\boldsymbol{w}}{\text{minimize}}
& & J(\boldsymbol{w}) \\
& \text{subject to}
& & 0<\rho_{b,j}^{\text{UL}}(w_b)<1, \; \forall b \in \mathcal{B}\notag\\
&
&&  0<\rho_{b,j}^{\text{DL}}(w_b)<1, \; \forall b \in \mathcal{B}.\notag
\end{align}
\section{Self-Organizing UL/DL Configuration}\label{sec:self_org}
To solve \Fref{eq:cost_prob}, we develop a distributed algorithm which dynamically optimizes the UL/DL configuration. The goal is to design a decentralized algorithm that selects a vector of switching points $\boldsymbol{w}$ that minimizes the cost function in \Fref{eq:cost_fun}. With the lack of global network information, the algorithm must rely only on the local information available at each SCBS to optimize an \emph{individual} cost function rather than the global cost in \Fref{eq:cost_fun}. However, the cost function for each SCBS depends not only on its own traffic load but also on the interference experienced from neighboring cells. Therefore, each SCBS $b$ should learn to estimate its cost function and use this estimated cost function to update its \emph{strategy}. Here, an SCBS's strategy is essentially the selection of a switching point.

In view of the interference coupling between neighboring cells, the performance of each SCBS depends not only on its choice of switching points, but on other SCBSs' choices as well. Therefore, we model this problem as a strategic noncooperative game $ \mathcal{G}=\Bigl(\mathcal{B},\{\mathcal{A}_b\}_{b\in\mathcal{B}},\{ J_b\}_{b\in\mathcal{B}} \Bigr)$ where $\mathcal{B}$ is the set of players (SCBSs), in which each of them selects its action $a_b^{(n_b)}$
from a set of actions $ \mathcal{A}_b  = \{a_b^{(1)},a_b^{(2)}, \ldots, a_b^{(N_b)}\}$, where $N_b$ is the number of possible actions, which corresponds to the number of switching points $W_f$ in our problem. For each BS $b\in \mathcal{B}$, the corresponding cost function can be expressed as follows:
\begin{multline}\label{eq:cost_fun2}
J_b(a_b^{(n_b)},\boldsymbol{a}_{-b}) =  \frac{1}{w_b}\sum_{j=1}^{w_b} \frac{\rho_{b,j}^{\text{UL}}}{1-\rho_{b,j}^{\text{UL}}} +\\
\frac{1}{W_f-w_b}\sum_{j=w_b+1}^{W_f} \frac{\rho_{b,j}^{\text{DL}}}{1-\rho_{b,j}^{\text{DL}}}
\end{multline}
where $a_b^{(n_b)}$ is the player's selected action and $\boldsymbol{a}_{-b}$ is the vector of other players' actions.

Each player $b$ chooses an action following a mixed strategy profile $\boldsymbol{\pi}_{b}=[\pi_{b,a_b^{(1)}},\pi_{b,a_b^{(2)}},\ldots,\pi_{b,a_b^{(N_b)}}]$, which is a vector of probability distributions over the set of possible actions $\mathcal{A}_b$. Let the strategy of choosing an action $a_b^{(n_b)}$ by player $b$ at a time frame $t$ be the probability that this action is selected $\pi_{b,a_b^{(n_b)}}(t) = \Pr(a_b(t)=a_b^{(n_b)})$. Then, by randomizing the action selection following their mixed-strategies, players aim at minimizing their long-term (expected) cost functions given by:
\begin{equation}\label{eq:expected_cost}
\bar{J}_b(\boldsymbol{\pi}_{b},\boldsymbol{\pi}_{-b})=\sum_{\boldsymbol{a}\in\mathcal{A}} J_b(a_b^{(n_b)},\boldsymbol{a}_{-b})\prod_{j=1}^{B}\pi_{j,a_j^{(n_j)}}
\end{equation}
where $\mathcal{A} = \mathcal{A}_1 \times \cdots \times \mathcal{A}_B$ is the space of action profiles.

In this game, each SCBS will choose the action that can lead to minimizing its cost function $J_b$, given other players' actions. We propose an algorithm that captures this behavior by adopting the Gibbs Sampling-based probability distribution, in which the probability of playing an action $a_b^{(n_b)}$ can be expressed as follows \cite{learning_journal}:
\begin{equation}\label{eq:gibbs}
\Lambda_{b,a_b^{(n_b)}} (\boldsymbol{a}_{-b}) = \frac{\exp\left(-\beta_b J_b(a_b^{(n_b)} ,\boldsymbol{a}_{-b})\right)} {\sum_{m=1}^{N_b}\exp\left(-\beta_b J_b(a_b^{(m)} ,\boldsymbol{a}_{-b})\right)}
\end{equation}
where $\beta_b$ is a Boltzmann's temperature coefficient. From \Fref{eq:gibbs}, it is clear that an action $a_b^{(n_b)}$ that yields a lower cost function will have a higher probability to be selected. Moreover, $\beta_b$ controls the exploitation versus exploration tradeoff, in which higher values lead to frequent selection of the actions with lower cost values, which is the exploitation case, while lower $\beta$ values lead to exploring other values as well.

Consequently, each player will run two coupled reinforcement learning processes to estimate its cost function and strategy vector. The goal of these processes is to find the strategies that allow achieving the best long-term performance while relying only on the instantaneous observations. These two processes run in parallel and allow each SCBS to build an estimate of its current cost function vector $\boldsymbol{\hat{J}}_b=[\hat{J}_{b,a_b^{(1)}},\ldots,\hat{J}_{b,a_b^{(N_b)}}]$ at time frame $t$ and use this estimate to update its current strategy profile vector $\boldsymbol{\pi}_b(t)$. These two processes can be written  $\forall b\in\mathcal{B}$ and $\forall n_b\in\{1,\ldots,N_b\}$ as follows:
\begin{align}\label{eq:cost_and_strategy_learn}
\begin{cases}
\hat{J}_{b,a_b^{(n_b)}}(t) =  \hat{J}_{b,a_b^{(n_b)}}(t-1)  + \\
 \hphantom{1cm}\alpha_b(t).\mathds{1}_{\{a_b(t-1)=a_b^{(n_b)}\}}\left(\tilde{J}(t-1) - \hat{J}_{b,a_b^{(n_b)}}(t-1) \right)\\
\pi_{b,a_b^{(n_b)}}(t) =  \pi_{b,a_b^{(n_b)}}(t-1) +\\
\hphantom{10}\hphantom{10}\hphantom{10}\hphantom{10}  \zeta_b(t).\left(\Lambda_{b,a_b^{(n_b)}} \bigl( \boldsymbol{\hat{J}}_b(t-1) \bigr) - \pi_{b,a_b^{(n_b)}}(t-1) \right)
\end{cases}
\end{align}
where $\tilde{J}(t-1)$ is the instantaneous observed cost function at time $t-1$, $\Lambda_{b,a_b^{(n_b)}}$ is given by \Fref{eq:gibbs},  $\alpha_b(t)$ and $\zeta_b(t)$ are the learning parameters, and should satisfy the following constraints \cite{learning_journal}:
\begin{align}\label{eq:converging_conditions}
\begin{cases}
   (i)\lim\limits_{T\rightarrow\infty}\sum\limits_{t=1}^{T}\alpha_b(t)=+\infty,\lim\limits_{T\rightarrow\infty}\sum\limits_{t=1}^{T}\alpha_b(t)^2<+\infty\\
   (ii)\lim\limits_{T\rightarrow\infty}\sum\limits_{t=1}^{T}\zeta_b(t)=+\infty,\lim\limits_{T\rightarrow\infty}\sum\limits_{t=1}^{T}\zeta_b(t)^2<+\infty\\
   (iii)\lim\limits_{t\rightarrow\infty}\frac{\zeta_b(t)}{\alpha_b(t)}=0.
\end{cases}
\end{align}
The proposed algorithm is illustrated in \Fref{algo:TDD_algorithm}.

\begin{algorithm}[t] 
 \begin{algorithmic}[1]

 \State \textbf{The implementation at each SCBS $b$}
 \State \textbf{Initialization}: pick a sequence of time frames $\{t_b^{(1)},t_b^{(2)},\ldots,t_b^{(n)},\ldots \}$, set $t_b^{(0)}=0$, $\boldsymbol{\hat{J}}_b(0)=(0,\ldots,0)$  and $\boldsymbol{\pi}_i(0)=\frac{1}{N_b}(1,\ldots,1)$.
 \For{each $t_b^{(n)}$}
 \State Select an action according to the probability distribution $\boldsymbol{\pi}_b(t_b^{(n-1)})$.
 \State Calculate the cell load according to \Fref{eq:load}.
 \State Calculate the observed cost function \Fref{eq:cost_fun2}.
 \State Update the estimated cost for the selected action and the probability distribution vector \Fref{eq:cost_and_strategy_learn}.
 \EndFor
 \end{algorithmic}
 \caption{Dynamic UL-DL Algorithm}
 \label{algo:TDD_algorithm}
\end{algorithm}
It is shown in \cite{learning_journal} that this reinforcement learning process guarantees that the algorithm convergences to the Logit Equilibrium (LE) \cite{logit_ref}. The LE is a special case of $\varepsilon$-Nash equilibrium in which none of the players can decrease its cost function by more than a value $\varepsilon$ without deviating from its current strategy. As $\varepsilon\rightarrow0$, the equilibrium coincides with the Nash equilibrium.

To explain the rationale behind the LE, we recall from \Fref{eq:gibbs} the effect of varying the Boltzmann's temperature coefficient $\beta_b$. As $\beta_b\rightarrow0$, the resulting mixed-strategy follows a uniform distribution, irrespective of the strategies of the other players, i.e., $\Lambda_{b,a_b^{(n_b)}} (\boldsymbol{a}_{-b}) = \frac{1}{N_b}$ for all $ a_b^{(n_b)}\in\mathcal{A}_b$. When $\beta_b\rightarrow\infty$, the result is a uniform distribution over the best actions given the strategies of the other players $\boldsymbol{\pi}_{-b}$.

For a finite $\beta_b>0$, higher probabilities are assigned to the actions associated with low average cost
and low probabilities to the actions associated with high cost values. Hence if a strategy profile $\boldsymbol{\pi}^{*}_b, \forall b\in\mathcal{B}$  provides the following bound for the cost reduction a player might obtain by unilaterally deviating from a given mixed-strategy \cite{learning_journal}:
\begin{equation}\label{eq:logit}
 \bar{J}_b(\boldsymbol{\pi}_b,\boldsymbol{\pi}_{-b}) - \bar{J}_b(\boldsymbol{\pi}'_b,\boldsymbol{\pi}_{-b}) \leqslant \frac{1}{\beta_b}\ln(N_b)
\end{equation}
where $\bar{J}_b(\boldsymbol{\pi}_b,\boldsymbol{\pi}_{-b})$ is the expected cost as defined in \Fref{eq:expected_cost}, then $\boldsymbol{\pi}^{*}_b$ is an LE equilibrium, i.e., an $\varepsilon$-equilibrium with $\varepsilon=\underset{b\in\mathcal{B}}\max\left(\frac{1}{\beta_b}\ln(N_b)\right)$.
This equilibrium highlights the tradeoff in choosing the value of the coefficient $\beta_b$. Although it follows from \Fref{eq:logit} that a reduction in cost obtained by a player unilaterally deviating from its strategy is more likely to occur using lower values of $\beta_b$, on the other hand, larger values make the $\varepsilon$-Nash equilibrium sufficiently close to the Nash equilibrium, as deduced from \Fref{eq:gibbs}.

\section{Simulation Results} \label{sec:sim_results}
In this section, we evaluate the performance of the proposed dynamic TDD algorithm. To illustrate the gains of the proposed scheme, we compare it against two baseline schemes; 1) \emph{fixed} TDD frame, in which small cells are assumed to have the same synchronous TDD frame, with equal UL and DL duty cycle, and 2) \emph{random} TDD frame, in which the switching point is varied randomly.

We consider an arbitrary number of SCBSs distributed randomly, and underlaying the macrocell  coverage area. In this work, we focus only on the SCBS-to-SCBS co-channel interference scenario. The bandwidth is assumed to be shared between all SCBSs. The bandwidth is assumed to be divided equally between all UEs transmitting or receiving in a given subframe. Both SCBSs and UEs are assumed to transmit with their maximum power, hence, no power control is considered in this work. We use the average packet throughput as the performance measure for different schemes, which is defined as the packet size divided by the delay encountered to complete its transmission. The motivation behind this is that it captures both packet rate and delay, which is the objective of the proposed scheme. To investigate the asymmetric UL/DL traffic, we conduct simulations for different mean UL-to-DL ratios. For example, UL-to-DL ratio of $0$ dB means that the average rate requirement $\lambda_b/\mu_b$ is the same for UL and
DL. Each SCBS uses a sequence of time frames, no more than a maximum of $200$ frames to learn its load and update its UL/DL configuration accordingly. The simulation parameters are summarized in \Fref{tab:parameters}.
\begin{table}[!t]
\renewcommand{\arraystretch}{1.3}
\caption{Simulation parameters}
\label{tab:parameters}
\centering
\begin{tabular}{p{1.4in}||p{1.4in}}
\hline
\bfseries Parameter & \bfseries Value/description\\
\hline\hline
System bandwidth & 10 MHz\\
\hline
Duplex mode & TDD\\
\hline
Number of SCBSs & [2,10]\\
\hline
Max. number of UEs per BS & 20\\
\hline
TDD frame length & 6 subframes\\
\hline
Sub-frame duration & 1 ms\\
\hline
Small cell radius & 40 m\\
\hline
Max. SCBS transmission power & 23 dBm\\
\hline
Max. UE transmission power & 23 dBm\\
\hline
Thermal noise & -174 dBm/Hz\\
\hline
Antenna configuration & 1*1\\
\hline
Path loss model & Multi-cell pico scenario \cite{standard_pathloss} \\
\hline
Penetration loss & 10 dB\\
\hline
Simulation time & 20 seconds (20000 subframes)\\
\hline
\end{tabular}
\begin{tabular}{p{3.2in}}
\centering \textbf{Learning parameters}
\end{tabular}
\begin{tabular}{p{1.4in}||p{1.4in}}
\hline
Strategy learning rate ($\zeta_b)$&$1/(t_b^{(n)})^{0.65}$\\
\hline
Load learning rate ($\alpha_b)$&$1/(t_b^{(n)})^{0.5}$\\
\hline
Temperature coefficient ($1/\beta_b$) &$0.005$\\
\hline
Maximum learning iterations & 200 frames\\
\hline
\end{tabular}
\end{table}

\begin{figure}
\centering
\includegraphics[width = 3.5in, trim=3.25cm .25cm 3.25cm .25cm, clip=true]{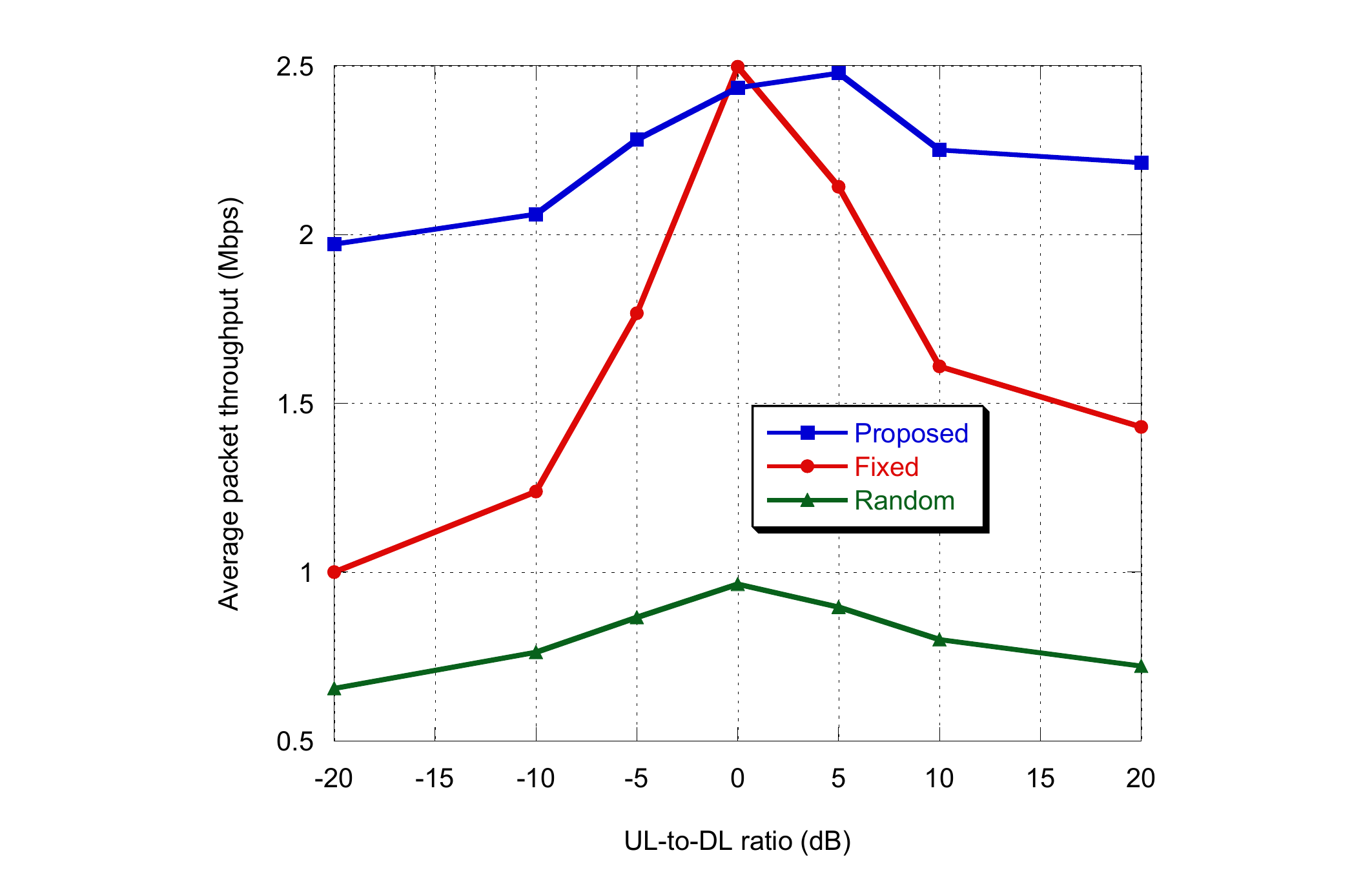}

\caption{Packet throughput performance in case of cells having the same UL-to-DL ratio for a network with 4 SCBSs.}
\label{fig:synch_perf}
\end{figure}

In \Fref{fig:synch_perf}, we compare the packet throughput performance of our scheme against the two baseline schemes for different UL-to-DL ratios. All SCBSs are assumed to have the same \emph{average} UL-to-DL ratios while the instantaneous traffic is different. The UL-to-DL ratio is expressed in dB, for example, $20$ dB means that $10\log(\frac{\lambda_b^{\text{UL}}/\mu_b^{\text{UL}}}{\lambda_b^{\text{DL}}/\mu_b^{\text{DL}}})=20$. \Fref{fig:synch_perf} shows that the proposed scheme achieves significant gains reaching up to $200\%$ at $-20$ dB compared to the random scheme in all traffic conditions. Moreover, this figure also shows that our approach outperforms the fixed scheme in case of asymmetric traffic conditions. The gain increases as the level of asymmetry increases, since the SCBSs are able to learn their UL and DL loads and adapt their transmissions accordingly. \Fref{fig:synch_perf} also shows that the proposed algorithm achieves up to $97\%$ gain at $-20$ dB over the fixed assignments. However,
the gain becomes smaller in the symmetric traffic case in which the fixed scheme is shown to achieve the same performance since it allocates equal resources to UL and DL and hence it is suitable for symmetric traffic.

\begin{figure}
\centering
\includegraphics[width = 3.5in, trim=3.25cm 0.25cm 3.25cm 0.25cm, clip=true]{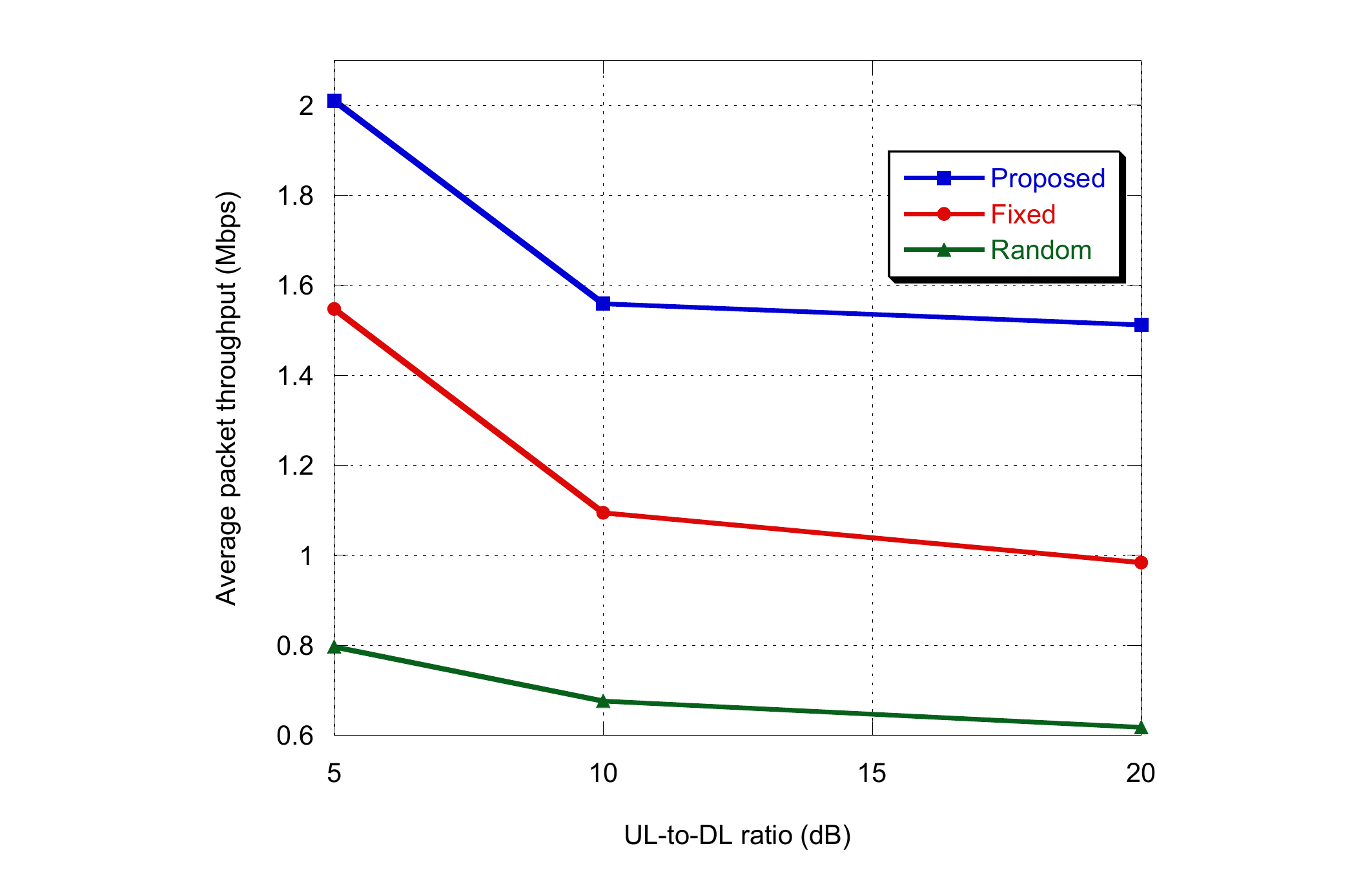}
\caption{Packet throughput performance in case of cells having opposite UL-to-DL ratios for a network with 4 SCBSs.}
\label{fig:asynch_perf}
\end{figure}

\Fref{fig:asynch_perf} shows the average packet throughput for the case in which half of the cells have opposite UL-to-DL ratios compared to the other half. For example, if the first half has a ratio of $10$ dB, the second half has a ratio of $-10$ dB. This scenario is challenging in the sense that it is associated with high cross-link interference. In \Fref{fig:asynch_perf}, we can see that the proposed scheme achieves considerable gains over both the random and fixed schemes. Clearly, the proposed algorithm is able to find a balance between selecting the switching point that matches the SCBS load and avoiding configurations that are associated with high cross-link interference. \Fref{fig:asynch_perf} shows that the proposed approach achieves gains reaching up to $145\%$ and $53\%$ over the random and fixed schemes, respectively in the case of UL-to-DL ratio of $20$ dB.

\begin{figure}
\centering
\includegraphics[width = 3.5in, trim=3.25cm .75cm 3.25cm .5cm, clip=true]{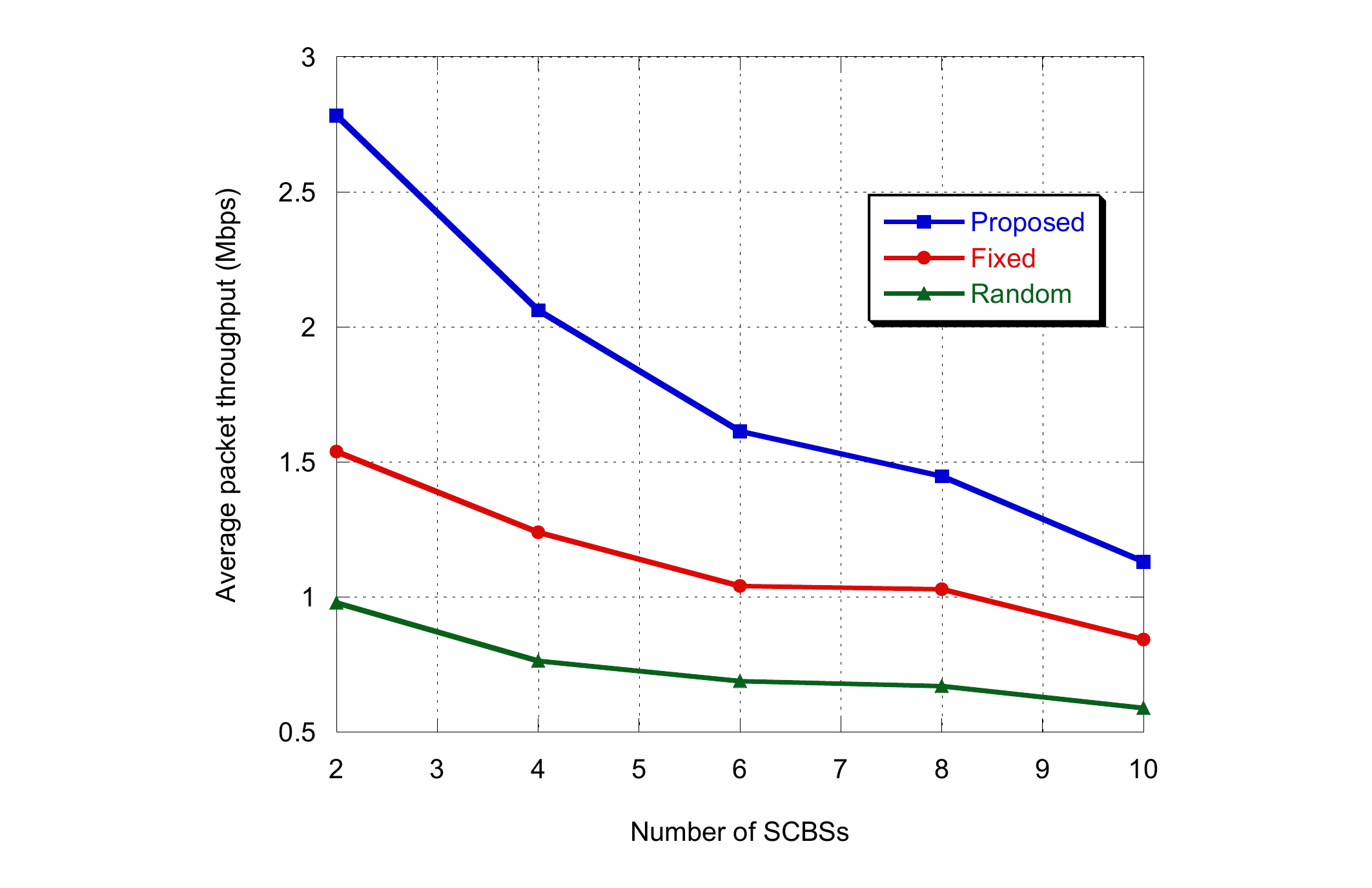}
\caption{Packet throughput performance at different network sizes.}
\label{fig:varying_num_cells}
\end{figure}

In \Fref{fig:varying_num_cells}, we show the average packet throughput resulting from all three schemes for different network sizes. This is done by varying the number of SCBSs while keeping the number of UEs per SCBS constant. All SCBSs are assumed to have an UL-to-DL ratio of $10$ dB. \Fref{fig:varying_num_cells} shows that, as the number of SCBSs increases, the average packet throughput resulting from all three schemes decreases. This is due to the fact that higher SCBSs density increases the effect of interference from neighboring SCBSs. However, \Fref{fig:varying_num_cells} shows that, for all network sizes, the proposed approach yields a higher average packet throughput than the baseline schemes. This performance gain reaches up to $90\%$ and $185\%$ relative to fixed and random schemes, respectively.

\begin{figure}
\centering
\includegraphics[width = 3.45in, trim=1.25cm 0.15cm 1.25cm 0.15cm, clip=true]{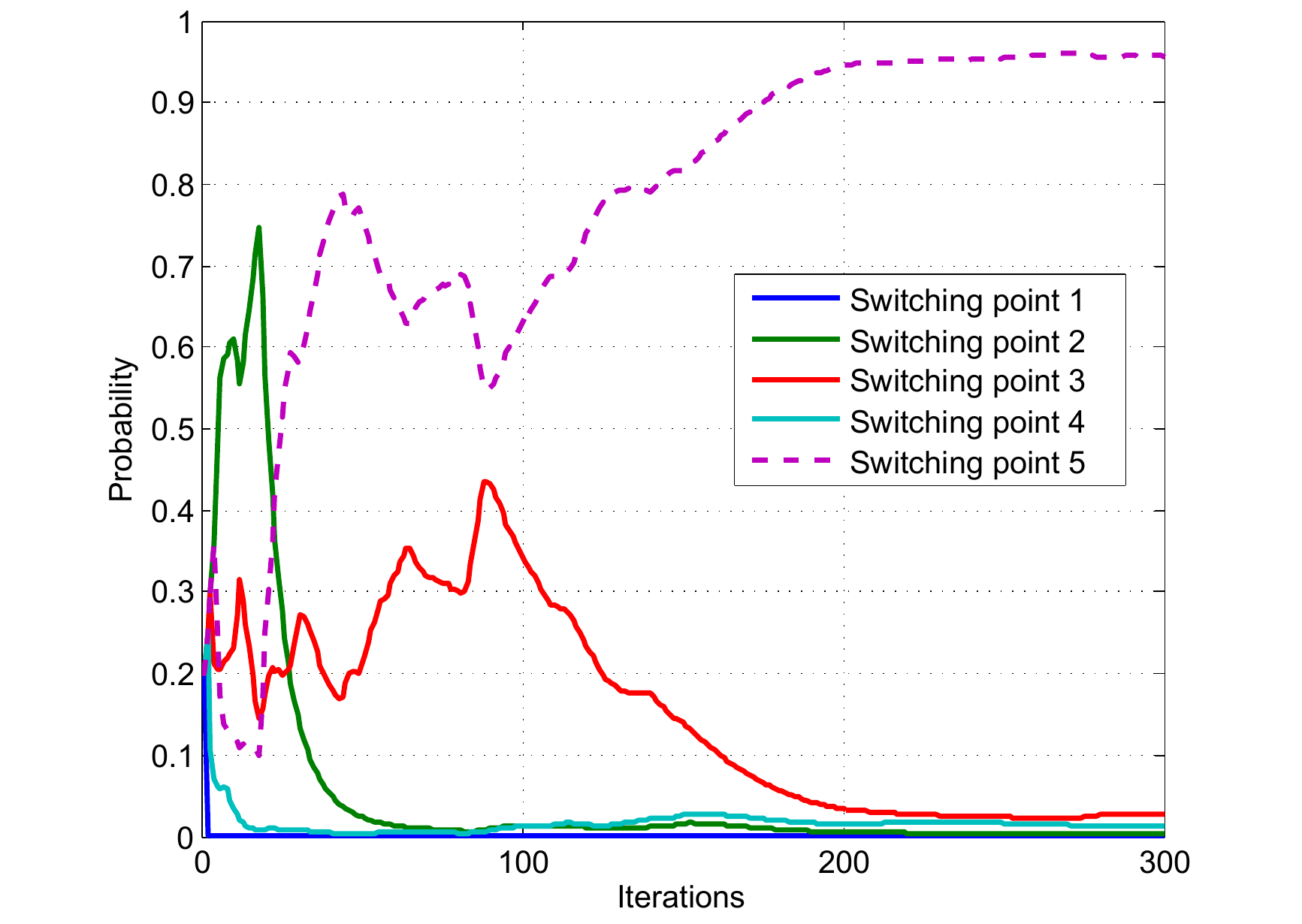}
\caption{Probability distribution of the actions set for the first SCBS.}
\label{fig:conv_cell1}
\end{figure}

\begin{figure}
\centering
\includegraphics[width = 3.45in, trim=1.25cm 0.15cm 1.25cm 0.2cm, clip=true]{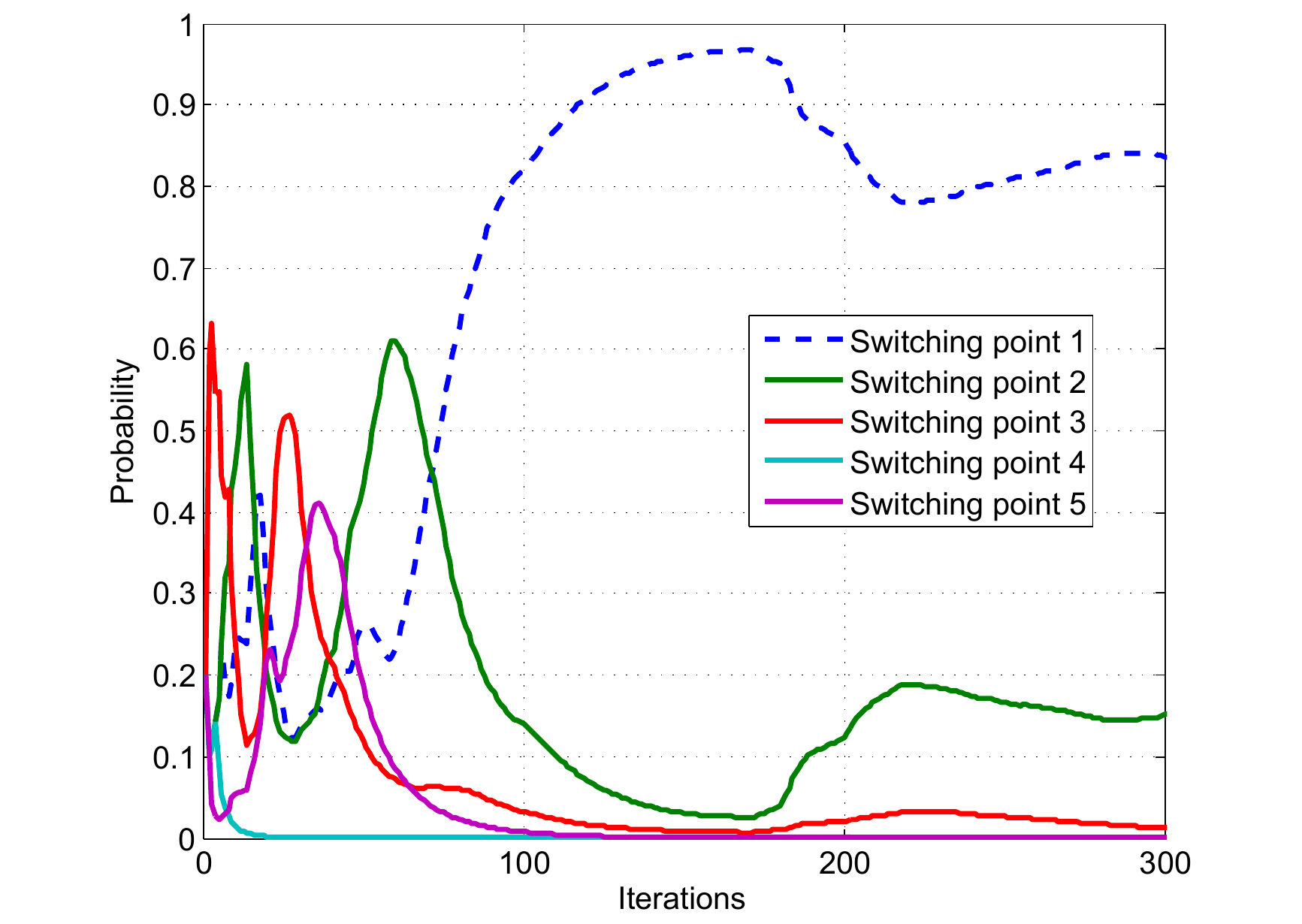}
\caption{Probability distribution of the actions set for the second SCBS.}
\label{fig:conv_cell2}
\end{figure}

In \Fref{fig:conv_cell1} and \Fref{fig:conv_cell2}, the convergence behavior of the proposed scheme is evaluated for an example of two SCBSs operating in opposite UL-to-DL ratios of $20$ dB and $-20$ dB. \Fref{fig:conv_cell1} shows the variations in the probability distribution for the set of actions (switching points) as the algorithm iterates in the first SCBS that is dominated by UL traffic, whereas the behavior of the second SCBS that is dominated by DL traffic is shown in \Fref{fig:conv_cell2}. From these figures, we can see that using the proposed algorithm, each SCBS is able to capture the traffic conditions autonomously while adapting its switching point to match the estimated traffic load. Switching point 5 which corresponds to the highest possible UL duty cycle, has the highest probability in the first cell, whereas in the second cell, switching point 1 is selected with the highest probability, which corresponds to the highest DL duty cycle. Interestingly, SCBSs are able to implicitly coordinate
their UL and DL transmissions, with no information exchange. From \Fref{fig:conv_cell1} and \Fref{fig:conv_cell2}, we can see that less than $200$ iterations are needed for the proposed algorithm to achieve convergence.\\

\section{Conclusions}\label{sec:conc}
In this paper, we have proposed a dynamic uplink and downlink configuration scheme that takes into account both the UL and DL loads as well as the interference from neighboring small cells. The proposed algorithm is distributed and relies only on the local observations to perform the UL and DL adaptation. Our results have shown that using the proposed algorithm, an SCBS is able to learn and estimate its current load then use it to optimize its strategy of selecting the proper UL/DL switching point. Simulation results have shown that the proposed approach significantly improves the network performance, in terms of the average packet throughput, compared to conventional fixed and random TDD deployments. Future work will investigate the problem of small cell clustering, power control, and the case in which the small cells adopt an open access policy.



%



\end{document}